\pdfoutput=0
\documentclass[prl,twocolumn,amsmath,amssymb,aps,superscriptaddress]{revtex4-1}
\usepackage{graphicx}
\usepackage{epstopdf}

\begin{document}

\title{Plasmon confinement in fractal quantum systems}

\address{School of Physics and Technology, Wuhan University, Wuhan 430072, China}
\address{Institute for Molecules and Materials, Radboud University, NL-6525 AJ Nijmegen, The Netherlands}

\author{Tom Westerhout}
\address{Institute for Molecules and Materials, Radboud University, NL-6525 AJ Nijmegen, The Netherlands}

\author{Edo van Veen}
\address{Institute for Molecules and Materials, Radboud University, NL-6525 AJ Nijmegen, The Netherlands}
\address{School of Physics and Technology, Wuhan University, Wuhan 430072, China}

\author{Mikhail I. Katsnelson}
\address{Institute for Molecules and Materials, Radboud University, NL-6525 AJ Nijmegen, The Netherlands}

\author{Shengjun Yuan}
\email{s.yuan@whu.edu.cn}
\address{School of Physics and Technology, Wuhan University, Wuhan 430072, China}
\address{Institute for Molecules and Materials, Radboud University, NL-6525 AJ Nijmegen, The Netherlands}

\date{\today}

\begin{abstract}
Recent progress in the fabrication of materials has made it possible to create
arbitrary non-periodic two-dimensional structures in the quantum plasmon regime.
This paves the way for exploring the plasmonic properties of electron gases in
complex geometries such as fractals. In this work, we study the plasmonic
properties of Sierpinski carpets and gaskets, two prototypical fractals with
different ramification, by fully calculating their dielectric functions. We show that the
Sierpinski carpet has a dispersion comparable to a square lattice, but the
Sierpinski gasket features highly localized plasmon modes with a flat
dispersion. This strong plasmon confinement in finitely ramified fractals
can provide a novel setting for manipulating light at the quantum scale.

\end{abstract}

\pacs{PACS}

\maketitle


Nowadays, different experimental techniques allow for the creation of arbitrary
non-periodic two-dimensional (2D) lattices. For example, artificial lattices
\cite{engineering2009, artificial2013}, systems consisting of quantum dots that
can be arranged in any custom shape, have attracted a lot of attention lately.
More generally, nanolithography methods can be used to make high-quality 2D
structures of arbitrary shape with a resolution in the order of tens of
nanometers \cite{scarabelli2015fabrication}. Other methods, such as molecular
self-assembly \cite{newkome2006nanoassembly, assembly2015} have been used to
grow Sierpinski gaskets. This presents an opportunity to experimentally study
complex 2D systems, such as fractals.

Fractals have no translational invariance, so where a Bloch description is
natural in the case of lattices, here it is not possible. Still, the
Schr\"odinger equation has been solved analytically on some simple fractals with
finite ramification \cite{kadanoff1983}. For others, like the Sierpinski carpet,
no analytical expressions for eigenenergies and eigenstates have been found yet.
The latter systems are better tackled numerically \cite{percolation1996}. It has
been shown that the quantum conductance of Sierpinski carpets exhibits fractal
fluctuations \cite{transport2016} and that their optical conductivity features
sharp peaks due to electronic state pairs at characteristic length scales
present within the carpet \cite{optics2017}. However, its plasmonic properties
have not been investigated yet.

Historically, in most plasmonic devices, the Fermi wavelength of the electrons
was much smaller than the plasmon wavelength which is of the order of the
geometric size of the system for standing waves. In other words, the
characteristic plasmon wave vector $q \ll k_F$, where $k_F$ is the Fermi wave
vector. In this regime, plasmons can be described classically and there is no
need to use a quantum mechanical approach \cite{nozieres1999theory,
platzman1973waves, vonsovskiui1989quantum, giuliani2005quantum}.  

Recently, due to the progress in nanodevice fabrication,
the quantum regime for plasmons has been reached \cite{scholl2012quantum,
tame2013quantum}. In this regime, localized surface plasmons make it possible to 
confine light to scales much smaller than the scales of conventional optics, and 
as such provide a unique way for light manipulation
on scales below the diffraction limit. Surface plasmons have found applications in 
surface-enhanced spectroscopy \cite{raman2005, second-harmonic1994}, 
biological and chemical sensing \cite{towards2005}, lithographic fabrication 
\cite{nanolithography2004}, and photonics \cite{brongersma2007surface}.

However, the theory of inhomogeneous quantum electron plasma,
even in the simplest random-phase approximation (RPA) \cite{nozieres1999theory,
platzman1973waves, vonsovskiui1989quantum, giuliani2005quantum}, is quite
complicated due to the essential nonlocality of the dielectric function
\cite{vonsovskiui1989quantum}. Recently, a rigorous scattering theory of
plasmons by obstacles was built \cite{torre2017lippmann}, but finding plasmon 
eigenmodes of inhomogeneous quantum systems still remains a challenge.
 As a matter of fact, this problem is very old, starting with the early considerations
\cite{bloch1933bremsvermogen, jensen1937eigenschwingungen} of ``atomic
plasmons'' \cite{kh1971ishmukhametov, sen1973spectrum, gadiyak1975collective,
ishmukhametov1975collective, amusia1978existence} which eventually turned out to
not exist \cite{verkhovtseva1976concerning, ishmukhametov1981existence}. 
Previous attempts use additional uncontrollable approximations such as
truncation of quantum states \cite{amusia1978existence}, semi-classical
\cite{kh1971ishmukhametov, ishmukhametov1975collective,
ishmukhametov1981existence} or even classical \cite{gadiyak1975collective}
approaches.

Here we will present the results of accurate, straightforward
calculations of plasmon spectra in an inhomogeneous quantum system with
nontrivial geometry, namely Sierpinski carpets and gaskets, two prototypic
examples of infinitely and finitely ramified fractals, respectively. These two
types of fractals can have widely different properties. For example, it has been
found that infinitely ramified fractals exhibit phase transitions not present in
finitely ramified fractals \cite{gefen1984phase}.

In this letter first we outline the methods used and present a numerical method
for calculation of plasmonic properties of systems with no translational
invariance that is applicable to arbitrary geometries. Then, we discuss the
results of these calculations on fractal systems. We compare the plasmon
dispersions of the Sierpinski carpet and gasket to those of a square and
triangle, respectively.



\begin{figure}[h]
    \input{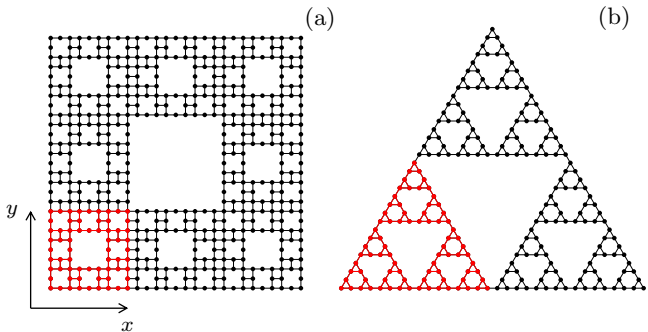}
    \caption{
        The fractals considered in this paper.
        (a) A third iteration Sierpinski carpet. The width of the sample is $3^3
        = 27$ lattice constants, or approximately $6.6\,\text{nm}$.
        (b) A fifth iteration Sierpinski gasket. Its width is $2^5 = 32$
        lattice constants ($7.9\,\text{nm}$).
        The previous iterations are indicated in red.
    }
    \label{fig:coordinates}
\end{figure}

We consider a system described by a tight-binding Hamiltonian
\begin{equation} \label{eq:hamiltonian}
    \hat H = - t \sum_{\langle a,b \rangle} \hat c^{\dagger}_{a \vphantom{b}}
    \hat c^{\vphantom{\dagger}}_b \;,
\end{equation}
where $t$ is the hopping parameter. Here, we have taken the on-site potential to
be zero and only consider nearest-neighbor hoppings. The two systems of interest
are illustrated in Fig. \ref{fig:coordinates}.

Fractals are made using an iterative process. For example, to make the
Sierpinski carpet, a previous iteration (indicated in red in Fig.
\ref{fig:coordinates}) is copied $\mathcal{N} = 8$ times to make a next
iteration that is $\mathcal{L} = 3$ times wider. With each fractal we can
associate a Hausdorff dimension, given by $d_H=\log_\mathcal{L}\mathcal{N}$, as
a measure for how space-filling its structure is. For the carpet $d_H \approx
1.89$, for the gasket $d_H \approx 1.58$.

Moreover, for each fractal we can define a ramification number, giving a measure
of how connected it is. The Sierpinski carpet is infinitely ramified: as a
higher iteration is taken, the number of bonds that need to be cut to separate
it from a lower iteration goes to infinity. In contrast, the Sierpinski gasket
is finitely ramified.

We use a hopping parameter $t=2.8\,\text{eV}$ and a lattice constant $a =
0.246\,\text{nm}$. These are the parameters for graphene, and they are
representative for 2D systems in general. Choosing a different lattice constant
will lead to a different plasmon spectrum, but the same qualitative behaviour.

Using this tight-binding model we obtain the exact eigenstates $|i\rangle$ with
corresponding eigenenergies $E_i$, to use for the calculation of the dielectric
function.

The dielectric function operator $\hat\varepsilon(\omega)$, by definition,
relates the external potential $\hat V_\text{ext}(\omega)$ to the total
potential $\hat V$:
\begin{equation} \label{eq:dielectric_function_def}
    \langle\mathbf{r}| \hat V_\text{ext}(\omega) |\mathbf{r}\rangle
        = \int\!\! \text{d}^d r' \; \langle\mathbf{r}| \hat\varepsilon(\omega)
            |\mathbf{r'}\rangle \langle\mathbf{r'}| \hat V |\mathbf{r'}\rangle
    \;.
\end{equation}
$d$ is the dimension of our problem. For the systems considered here $d = 2$.
Treating $\hat V$ as a perturbation, within RPA, the dielectric function may be
expressed as follows \cite{vonsovskiui1989quantum}:
\begin{equation} \label{eq:diel_func}
\begin{aligned}
    \langle\mathbf{r}| \hat\varepsilon(\omega) |\mathbf{r'}\rangle
        &= \langle\mathbf{r}|\mathbf{r'}\rangle
        \!-\!\!\!
        \int\!\! \text{d}^d r'' \langle\mathbf{r}| \hat V_\text{C}
            |\mathbf{r''}\rangle \langle\mathbf{r''}| \hat\chi(\omega)
            |\mathbf{r'}\rangle \;, \\
    \langle\mathbf{r}| \hat V_\text{C} |\mathbf{r''}\rangle
        &\equiv \frac{e^2}{\|\mathbf{r} - \mathbf{r''}\|} \;, \\
    \langle\mathbf{r''}| \hat\chi(\omega) |\mathbf{r'}\rangle
    &= g_s \cdot\! \lim_{\eta \to 0+} \sum_{i,j}
            \langle i| \hat G |j\rangle
            \langle j|\mathbf{r''}\rangle \langle\mathbf{r''} |i\rangle
            \langle i|\mathbf{r'}\rangle \langle\mathbf{r'} |j\rangle \;, \\
    \langle i| \hat G |j\rangle
        &\equiv \frac{n_i - n_j}{E_i - E_j - \hbar(\omega + i\eta)} \;.
\end{aligned}
\end{equation}
$|\mathbf{r}\rangle$
denotes a position eigenvector; $\hat V_\text{C}$ is the Coulomb interaction
potential; $\hat\chi(\omega)$ is the polarizability function; $\eta$ is the
inverse relaxation time; $g_s = 2$ is spin degeneracy; $n_i$ is $i$'th energy
level occupational number according to the Fermi-Dirac distribution
\begin{equation} \label{eq:FD_dist}
    n_i = \frac{1}{e^{(E_i - \mu)/kT} + 1}.
\end{equation}
We used room temperature $T=300\,\text{K}$ and an inverse relaxation time $\eta
= 6\,\text{meV/}\hbar$.

Eqs. \eqref{eq:diel_func} allow us to exactly calculate the full dielectric
function $\hat\varepsilon(\omega)$ of any tight-binding system without
translational invariance. The open source project documentation
\cite{Westerhout2017} lists the computational techniques employed which, despite
the $\mathcal{O}(N^4)$ algorithmic complexity, make calculations possible for
systems of up to several thousands of sites.

To visualise the plasmon modes in a quantum mechanical system Wang et al
\cite{plasmonic2015} introduced the following method. Consider the dielectric
function in its spectral decomposition:
\begin{equation}
    \hat\varepsilon(\omega) = \sum_n \epsilon_n(\omega)|\phi_n(\omega)\rangle
        \;.
\end{equation}
In this method, for each $\omega$ we consider only the eigenvalue
$\epsilon_{n_1(\omega)}(\omega)$ that has the highest value of
$\;-\operatorname{Im}[1/\epsilon_n(\omega)]$, which gives us the plasmon
eigenmode $|\phi_{n_1(\omega)}(\omega)\rangle$ that contributes most to the loss
function.

However, it is not clear how to access these plasmon modes experimentally.
Currently, the standard way of probing plasmon properties of small quantum
mechanical systems is EELS. The fact that we calculate the \emph{full}
dielectric function gives us the possibility to calculate the following Fourier
transform, which distinguishes this study from others:
\begin{equation}
    \langle\mathbf{q} |\hat\varepsilon(\omega)| \mathbf{q}\rangle
        = \frac{1}{(2\pi)^d} \int\!\!\text{d}^d r \!\!\!\int\!\!\text{d}^d r'\;
            \langle\mathbf{r} |\hat\varepsilon(\omega)| \mathbf{r'}\rangle\,
                e^{-i\mathbf{q}(\mathbf{r} - \mathbf{r'})} \; .
\end{equation}
The loss function $\; -\operatorname{Im}[1 /
\langle\mathbf{q}|\hat\varepsilon(\omega)| \mathbf{q}\rangle]$ is then directly
measurable using EELS techniques \cite{nozieres1999theory, platzman1973waves,
vonsovskiui1989quantum, giuliani2005quantum, lu2009plasmon}.


\begin{figure}[h]
    \input{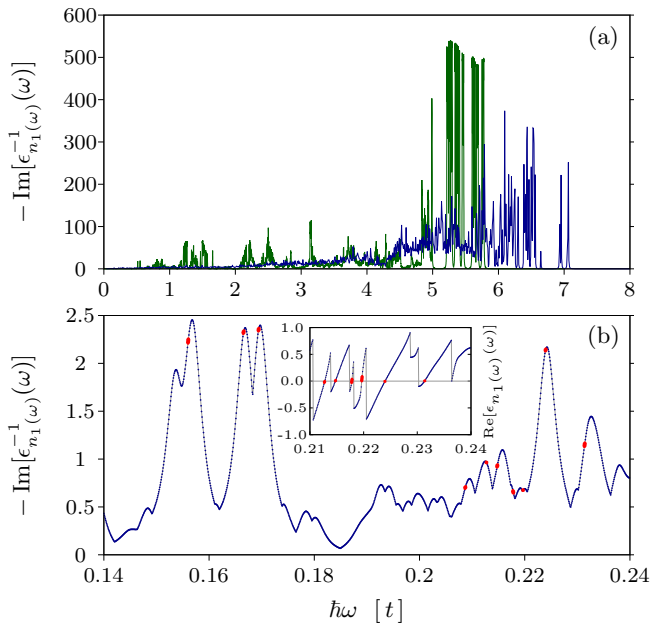}
    \caption{
        The highest contribution to the loss function
        $\;-\operatorname{Im}[\epsilon^{-1}_{n_1(\omega)} (\omega)]$. (a) The
        loss function for the entire range of frequencies, in the case of (blue)
        a third iteration  Sierpinski carpet and (green) a sixth iteration
        Sierpinski gasket. (b) The loss function of a third iteration
        Sierpinski carpet for a frequency interval $0.14t < \hbar\omega <
        0.24t$. Inset: $\operatorname{Re}[\epsilon_{n_1(\omega)} (\omega)]$ for
        a frequency interval $0.21t < \hbar\omega < 0.24t$, showing
        discontinuities. Red dots indicate pairs of points between which
        $\operatorname{Re}[\epsilon_{n_1(\omega)}(\omega)]$ crosses zero in a
        continuous manner.
    }
    \label{fig:spectrum}
\end{figure}

Formally, there are two ways of identifying plasmons. A plasmon frequency is
either given by a local maximum of the loss function
$\;-\operatorname{Im}[1/\epsilon_{n_1(\omega)}(\omega)]$, or by a frequency at
which $\operatorname{Re}[\epsilon_{n_1(\omega)}(\omega)] = 0$. These frequencies
are not exactly equal due to Landau damping, which is quantified by $\eta$
\cite{andersen2012spatially}.

The real-space loss function of the highest contributing plasmon mode is shown
in Fig. \ref{fig:spectrum}. It shows that there is a large number of plasmon
frequencies, and that the associated losses increase with increasing frequency.
At each discontinuity in $\operatorname{Re}[\epsilon_{n_1(\omega)}(\omega)]$ a
different mode is found to be the highest contributor to the loss function. Such
a discontinuity is not associated with a plasmon, even though
$\operatorname{Re}[\epsilon_{n_1(\omega)}(\omega)]$ switches sign.

\begin{figure}[h]
    \includegraphics[width=8cm]{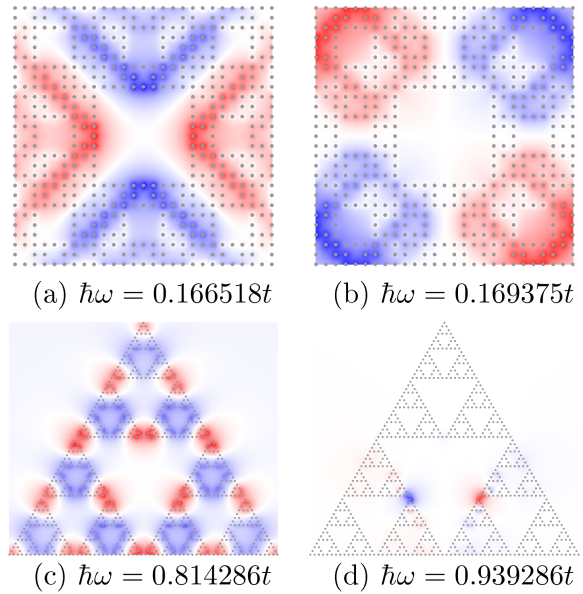}
    \caption{
        The highest contributing plasmon eigenmodes in real space. A few
        examples of the real space distribution
        $\operatorname{Re}[\langle\mathbf{r}|
        \phi_{n_1(\omega)}(\omega)\rangle]$ of plasmon modes for (a,b) a third
        iteration Sierpinski carpet and (c,d) a sixth iteration Sierpinski
        gasket. Eigenmodes exhibiting different characteristic length scales are
        shown.
    }
    \label{fig:eigenstates}
\end{figure}

The real part of the highest contributing plasmon eigenmodes for both the carpet
and gasket are shown in Fig. \ref{fig:eigenstates}. For further analysis, the
inverse participation ratio $\operatorname{IPR}(\omega) = \int\!\!\text{d}^d r
|\langle \mathbf{r}|\phi_{n_1(\omega)} \rangle|^4$ can give us a measure of
localization. The average IPR of $|\phi_{n_1(\omega)}\rangle$ was found to be an
order of magnitude higher for the gasket than for the carpet. This can be seen
as a consequence of the finite ramification of the gasket, i.e. the fact that it
is less connected, and therefore the electrons are more confined and exhibit
more localized plasmon eigenmodes. Fig. \ref{fig:eigenstates}(d) shows an
example of such a highly localized mode.

\begin{figure*}[t]
    \includegraphics{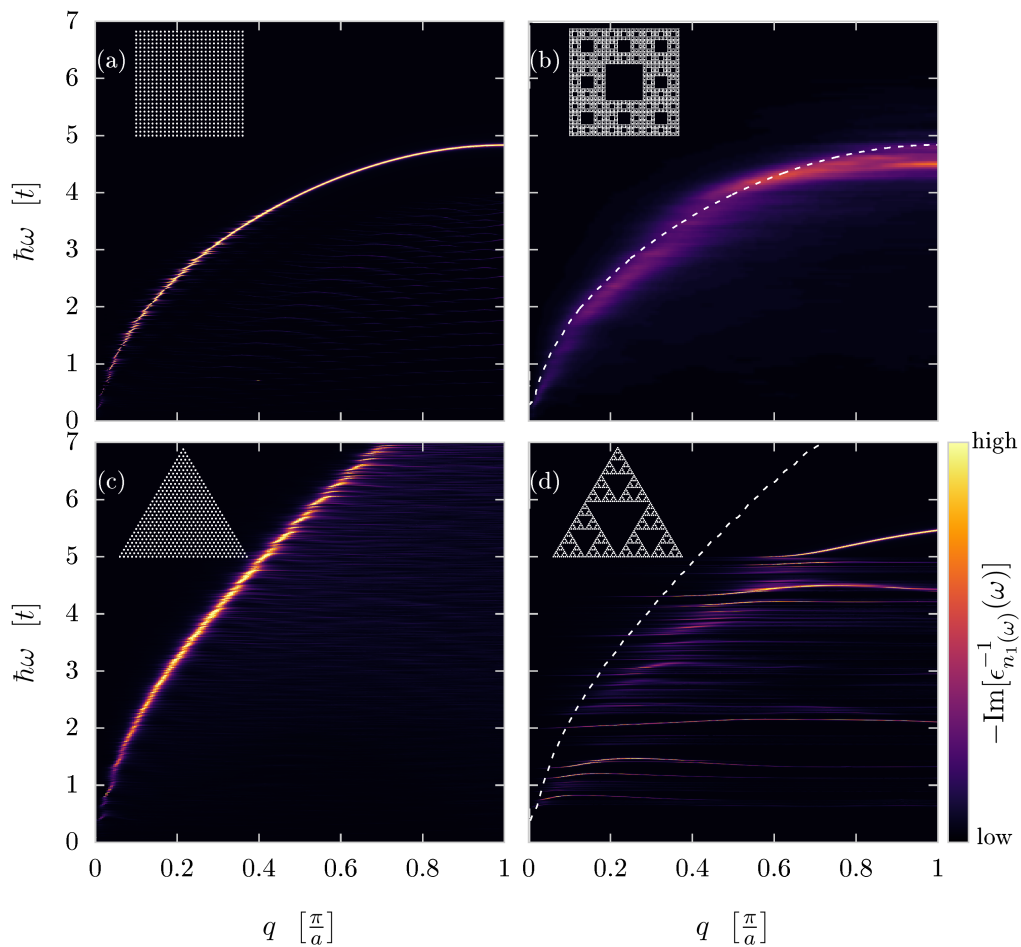}
    \caption{
        Dispersion relation $\;-\operatorname{Im}[1 / \langle\mathbf{q}|
        \hat\varepsilon(\omega) |\mathbf{q}\rangle]$, showing the frequency and
        momentum dependency of the loss function. Momentum was taken along the
        $x$--axis. (a) A square built out of square lattice as compared to (b)
        the fourth iteration Sierpinski carpet. Similarly, (c) a triangle built
        out of triangular lattice as compared to (d) a sixth iteration
        Sierpinski gasket. The maximum of the left hand side is plotted as a
        dashed white line on the right hand side.
    }
    \label{fig:dispersion}
\end{figure*}

We now turn to the Fourier transform of the real-space loss function in order to
make a comparison to EELS experiments. Fig. \ref{fig:dispersion} shows the loss
function as function of both $q$ and $\omega$.

There is a close resemblance between the carpet (Fig. \ref{fig:dispersion}(a))
and a square sample (Fig. \ref{fig:dispersion}(b)). The dispersion of the carpet
has extra broadening, similar to the broadening found in systems with disorder
\cite{jin2015screening}. However, generally speaking, both curves look like a
regular $\varepsilon(\omega) \propto \sqrt{q}$ dispersion relation for surface
plasmons \cite{giuliani2005quantum}. The carpet exhibits no translational
invariance, i.e. $q$ is not actually a good quantum number, so this behavior is
quite remarkable. The dispersion of the fourth iteration Sierpinski carpet is
already very close to the third iteration dispersion. This convergence indicates 
that the result is representative for the real fractal at infinite iteration.

For the Sierpinski gasket (Fig. \ref{fig:dispersion}(c)), we observe different
behavior. This fractal does not closely follow the dispersion relation of a
triangle built out of a triangular lattice (Fig. \ref{fig:dispersion}(d)).
Instead, we can clearly see the formation of multiple localized modes with near
flat dispersion. Again, this result is reasonably converged.


Concluding, in this work we have calculated the plasmon dispersion for the
Sierpinski carpet and Sierpinski gasket. The Sierpinski carpet has a plasmon
dispersion comparable to the dispersion of a square lattice, whereas the gasket
exhibits highly localized plasmon modes. More generally, a finitely ramified fractal 
can exhibit strong plasmon confinement, providing a novel setting
for the manipulation of light at the quantum scale. With current experimental 
techniques, these results can be probed experimentally.
Moreover, we have presented a rigorous approach for calculating plasmonic
properties of generic tight-binding systems, published as an open source
software project \cite{Westerhout2017}. We believe that this code can be very
useful for future projects relating to plasmonic properties of
non-translationally invariant systems.


This work was supported by the National Science Foundation of China under Grant
No.  11774269 and by the Dutch Science Foundation NWO/FOM under grant No.
16PR1024 (S.Y.), and by the European Research Council Advanced Grant program
(contract 338957) (M.I.K.). Support by the Netherlands National Computing
Facilities foundation (NCF), with funding from the Netherlands Organisation for
Scientific Research (NWO), is gratefully acknowledged.

\bibliography{bibliography}

\end{document}